\documentclass{article}
\usepackage{PRIMEarxiv}

\usepackage[utf8]{inputenc} 
\usepackage[T1]{fontenc}    
\usepackage{hyperref}       
\usepackage{url}            
\usepackage{booktabs}       
\usepackage{amsfonts}       
\usepackage{nicefrac}       
\usepackage{microtype}      
\usepackage{lipsum}
\usepackage{fancyhdr}       
\usepackage{graphicx}       
\graphicspath{{media/}}     
\usepackage{natbib}
\usepackage{amsmath} 
\usepackage{multirow} 
\usepackage{amssymb} 
\usepackage{natbib}
\setcitestyle{numbers,square}

\title{Solar Radio Burst Detection Based on Deformable DETR
\thanks{\textit{\underline{Citation}}: 
\textbf{Mingming Wang et al. Solar Radio Burst Detection Based on Deformable DETR. Accepted for publication in The Astrophysical Journal Supplement.}} 
}

\author{
  Mingming Wang \\
  School of Information Science and Engineering \\
  Yunnan University\\
  Kunming, China\\
   \And
  Guowu Yuan* \\
 School of Information Science and Engineering \\
  Yunnan University\\
  Kunming, China\\
  \texttt{gwyuan@ynu.edu.cn} \\
   \And
 Hao Zhou \\
 School of Information Science and Engineering \\
  Yunnan University\\
  Kunming, China\\
   \And
  Chengming Tan \\
 State Key Laboratory of Space Weather \\
  National Space Science Center\\
  Beijing, China\\
   \And
 Hao Wu  \\
 School of Information Science and Engineering \\
  Yunnan University\\
  Kunming, China\\
  }

\begin{document}

\maketitle

\begin{abstract}
 Solar radio bursts (SRBs) detection is crucial for solar physics research and space weather forecasting. The main challenges faced are noise interference in the spectrum and the diversity of SRBs. However, most research focuses on classifying whether SRBs exist or detecting a single type of SRBs. Existing detection models exhibit deficiencies in the accuracy of SRBs detection. Moreover, existing detection models cannot effectively handle background noise interference in solar radio spectrograms and the significant scale variations among different burst types. This paper proposes a high-performance detection model for solar radio bursts (SRBs) based on Deformable DETR (DEtection TRansformers) called DETR4SBRs. Firstly, this study designed a scale sensitive attention (SSA) module better to address the scale variations of SRBs. Subsequently, this study introduced collaborative hybrid auxiliary training to mitigate the positive-negative sample imbalance issue in Deformable DETR. The experimental results demonstrate that the proposed model achieves a mAP@50 of 83.5$\%$ and a recall rate of 99.4$\%$ on the SRBs dataset. Additionally, the model exhibits excellent noise-robust performance and can efficiently detect and locate Type II, III, IV, and V SRBs. The model proposed in this study provides robust support for preliminary solar radio burst data processing and has significant implications for space weather forecasting. The source code and data are available on the \href{https://github.com/onewangqianqian/SSA-Co-Deformable-DETR.git}{https://github.com/onewangqianqian/SSA-Co-Deformable-DETR.git} and archived on Zenodo.
\end{abstract}


\section{Introduction}

Solar radio bursts (SRBs) are notable events in solar activity, marked by a sudden increase in solar radiation in the radio wavelength band. These SRBs are usually closely linked to the Sun's magnetic activity, especially events like Coronal Mass Ejections (CMEs) and solar flares \citep{marque2018solar}. They represent the evolution and interaction of the background plasma, non-thermal electrons, and magnetic fields in active regions. These SRBs are essential for diagnosing the solar corona's magnetic field and other aspects of the solar atmosphere\citep{young2018solar}. Solar radio bursts release a significant amount of electromagnetic radiation. This radiation impacts the space environment around Earth and interferes with communication systems and satellite operations.  \citep{Sato2019Solar}. The high-energy particle flux generated by these bursts poses radiation risks to spacecraft and astronauts, threatening aviation safety \citep{Yue2018The}. Intense solar radio activity can trigger geomagnetic storms, leading to ground electrical currents' fluctuations. These fluctuations may damage electrical infrastructure and cause power outages \citep{Muhammad2015Performance}.

Detecting SRBs is crucial in solar physics research and space weather forecasting. Detailed detection and classification of solar radio spectrograms form the data foundation for this research \citep{singh2019automated}. Various observatories and research institutions globally use solar radio spectrometers to monitor solar bursts around the clock across different frequency ranges, gathering vast amounts of observational data. Analysis of e-CALLISTO observational data shows that solar radio bursts' cumulative duration is only about 0.3$\%$ of the total observation time. These valuable and effective times of solar radio bursts are crucial for research in space weather forecasting. Traditional manual detection methods for solar radio bursts primarily rely on human observation and classification. These methods require experts to visually inspect solar radio spectrograms in order to identify and categorize different types of radio bursts. This approach is not only time-consuming but also prone to human error. This method consumes significant human resources, is inefficient, and is prone to errors. Therefore, automating the detection and classification of SRBs would greatly advance solar physics research and space weather forecasting.

With the development of astronomical observation equipment, astronomers now face the challenge of processing and analyzing unprecedented amounts of data. Deep learning technology in astronomical data processing has proliferated, becoming a powerful tool for handling and analyzing large datasets \citep{kremer2017big}. Recently, researchers have started using deep learning techniques to classify and detect solar radio spectrum.

\section{Related Works}
Early detection methods for solar radio bursts (SRBs) primarily depended on manual selection. Researchers examined solar radio spectrum captured by radio telescopes to recognize SRBs \citep{Zhu2020Supervised}.

Currently, many studies primarily focus on the classification of SRBs. Research has gradually progressed to the multi-class classification of SRBs, starting with simple binary classification (i.e., burst vs. non-burst). The work of \citep{ma2017multimodal} uses a multi-modal network to set radio signals of different characteristics and frequency bands as different modes. This approach successfully classified bursts and non-bursts, achieving an average classification accuracy of over 75$\%$. The study in \citep{li2022self} proposed a self-supervised learning method for classifying only burst and non-burst events in solar radio spectrograms. The issue of poor transfer learning performance was resolved by utilizing self-supervised training and self-masking techniques from natural language processing. This method overcomes the challenges posed by the significant differences between natural images and solar radio spectrogram images. It achieved classification accuracy similar to that of supervised learning. The research in \citep{yu2017solar} attempted to classify solar radio spectrograms using Long Short-Term Memory (LSTM) networks. This method captures features during the slow variation process of solar radio emissions, achieving high classification accuracy. The study in \citep{guo2020auto} proposes a hybrid neural network model that combines convolutional and memory units. This model effectively extracts the frequency structure features and time series characteristics of solar radio spectrogram images. This approach significantly improved the average classification accuracy of solar radio spectrograms to 98.73$\%$. In the study by \citep{chen2022classification}, a solar radio spectrogram classification method based on the Swin Transform was proposed. By transferring pre-trained model parameters and freezing hidden layer weights, this method effectively addressed the issues of sparse and uneven SRB spectrogram data. It also classified radio spectrogram bursts and non-bursts using relatively few model parameters. The research in \citep{scully2023improved} utilized Generative Adversarial Networks (GANs) to generate simulated Type III SRB data. Combining observations from the Low-Frequency Array (LOFAR) \citep{Haarlem2013LOFAR:}, the YOLOv2 object detection model was trained. This approach automatically detected and classified Type III SRBs, reaching a mAP of 77.71$\%$. The study in \citep{zhang2021auto} used solar radio spectrogram data from the Culgoora and Learmonth observatories. They proposed a conditional information-based Deep Convolutional GAN (C-DCGAN) model to classify five types of SRBs events automatically. This method partially resolved the overfitting problem caused by insufficient data samples.

With the rapid development of object detection models in deep learning, researchers have also begun to focus on detecting SRBs. In \citep{zhang2020auto}, Faster R-CNN was used to identify Type III bursts and small-scale spike burst events. It then extracted features such as starting and ending frequencies. Additionally, it extracted frequency drift rates and positional coordinates. In \citep{guo2022deep}, radio burst spectrogram data collected by the Green Bank Solar Radio Burst Spectrometer (GBSRBS) \citep{Magdalenić2012FLARE-GENERATED} were utilized. The study accomplished the recognition and localization of multiple types (Type II, Type III, and Type V) of SRBs events based on deep-learning object detection algorithms. In \citep{he2023solar}, an automatic recognition and localization method for SRB events based on a lightweight object detection model was proposed. Using e-CALLISTO observational data, a dataset containing Type II, III, IV, and V SRBs was established. The study improved the Vision Transformer with a self-attention mechanism and employed a lightweight model for detection. The proposed method achieved an average precision of 78.2$\%$ and a recall rate of 92$\%$ on the established SRB dataset.

Current methods for detecting SRBs face several issues: (1) they are limited to detecting only one or a few types of SRBs, resulting in a narrow range of detectable types; (2) due to background noise interference and the diverse scales of SRBs, the detection accuracy is not high.

\section{Contribution}
Given the aforementioned issues with SRB detection, an accurate multi-type SRB detection model must be proposed. The DETR \citep{carion2020end} (DEtection TRansformer) model is an end-to-end object detection framework. The DETR model has achieved success in many areas. In medical image processing, the capabilities of DETR are applied to tumor detection and the identification of other pathological features \citep{Xu2024Understanding}. In autonomous driving, DETR can be utilized for real-time traffic signs, pedestrians, and vehicle detection \citep{Zhao2019Detection}. In industrial production lines, DETR is applied to detect defects in the appearance of cigarettes \citep{Ding2025ESFDETR}. Additionally, DETR effectively distinguishes between complex objects and land use types when processing remote sensing images \citep{Han2023Capsule-inferenced}, making it widely applicable in environmental monitoring and urban planning. Traditional object detection methods \citep{girshick2015fast} typically require predefined anchors and candidate regions for prediction. The DETR employs an end-to-end structure to directly output a set of object predictions from the input image. This includes their classes and bounding boxes, without any preprocessing or postprocessing step. The convolutional neural network (CNN) is a class of neural networks designed specifically for processing image data. The convolutional kernel (also known as the filter) is a key component of the convolutional neural network. A convolutional kernel is a matrix used in convolutional neural networks to extract features. It slides over the input image and performs a dot product operation to generate feature maps. The receptive field refers to the region's size in the input image that influences a specific location in a feature map. CNN-based networks \citep{huang2018convolutional} use convolutional kernels with fixed receptive fields that can only capture local features of fixed sizes. The DETR-based model uses the encoder-decoder architecture of the Transformer \citep{vaswani2017attention} to process images. This allows the model to capture global contextual information effectively. Additionally, it can handle complex relationships between different features. These advantages of the DETR-based model make it suitable for the automatic processing of SRBs. Currently, there is no related research work in this area.

In object detection, a target refers to a specific object that needs to be identified and localized within an image. In this study, the target is solar radio bursts. Deformable DETR is a DETR model that is friendly to small objects and exhibits excellent training efficiency. To address the complexity of SRBs, we improved Deformable DETR to detect SRBs. The attention mechanism \citep{bahdanau2014neural, vaswani2017attention} simulates human visual focus. Its purpose is to improve the efficiency and effectiveness of neural networks in processing information. It enables the model to selectively focus on important parts of the input data while ignoring irrelevant information. This ability enhances the model's capacity to capture contextual information. We propose a novel attention mechanism—Scale Sensitive Attention (SSA) (full details in Section 5.1). This mechanism introduces dynamic convolution into the attention structure, replacing the deformable attention mechanism in the original Deformable DETR. Next, we employed collaborative hybrid auxiliary training \citep{zong2023detrs}. Collaborative hybrid auxiliary training is a training method designed for the DETR-base model. Collaborative hybrid auxiliary training uses traditional detection models as auxiliary heads to update the transformer decoder. It addresses the positive-negative sample imbalance issue in Deformable DETR (full details in Section 5.2). Finally, we adopted the Generalized Intersection over Union (GIoU) loss function \citep{rezatofighi2019generalized}  (full details in Section 5.3). GIoU is a metric used to evaluate the accuracy of bounding box predictions in object detection models. This function more precisely guides the model in adjusting bounding boxes. As a result, it improves localization accuracy and enhances the model's adaptability to complex scenes. Experimental results show that the proposed model can accurately detect Type II, III, IV, and V SRBs.

The main contributions of this paper are summarized as follows:
\begin{enumerate}
    \item	The DETR4SBRs model proposed in this paper can handle the complex scale variations and noise interference in SRBs. It demonstrates high performance.
    \item	This paper introduces the scale-sensitive attention (SSA) mechanism to address the shortcomings of the deformable attention mechanism in dealing with the complex scale issues of SRBs. By incorporating dynamic convolution into the attention structure, this mechanism performs well in continuously detecting SRBs.
    \item	The model in this paper achieves end-to-end object detection without any manually designed components. This simplifies the model's structure and training process while reducing the model's dependency on specific prior knowledge. It also demonstrates the strong generalization ability of the DETR4SBRs model.
\end{enumerate}

\section{SRBs Dataset}
\begin{figure}[htb]
    \centering
    \includegraphics[width=1.0\linewidth]{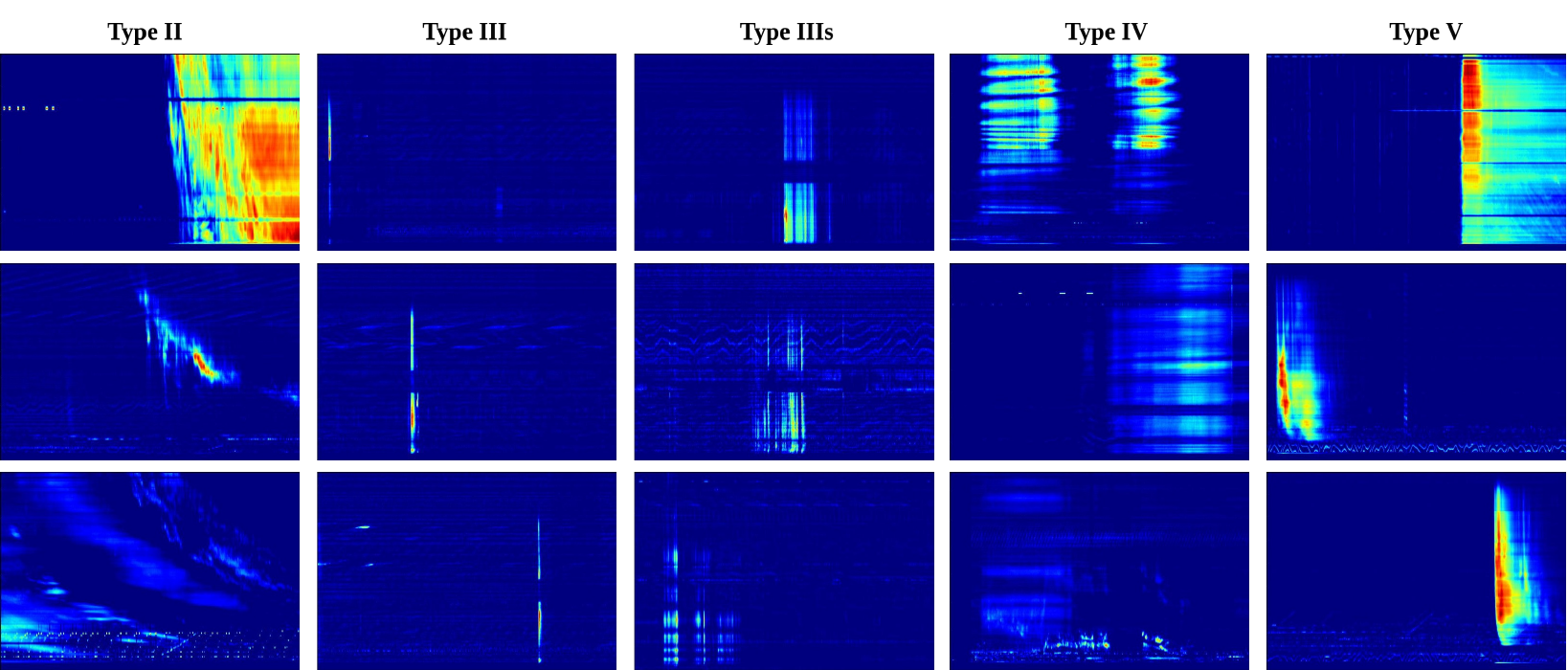}
    \caption{Illustrations of the five types of SRBs (where Type IIIs represents Type III SRBs appearing in group form).}\label{fig:fig1}
\end{figure}

Common SRBs can be classified into five different types, as shown in Figure \ref{fig:fig1}. Type I \citep{Sy1973A} SRBs are characterized by noise storms or continuous radiation. They manifest as a steady enhancement of background radio emission over a wide frequency band. Type I SRBs usually accompany Type IV SRBs and can be intense, covering more than 300 MHz. They may occur during and after major solar events such as flares and can last 10 to 300 minutes. Type II \citep{Gopalswamy2013Type} SRBs are slow-drift bursts that occur after large solar flares. They exhibit strong, narrowband radiation that sometimes drifts slowly and irregularly towards lower frequencies. Type III SRBs are characterized by rapid frequency drift, with typical durations of about one second. At lower frequencies,  the duration of Type III \citep{Reid2014A}  SRBs increases, and their frequency drift rate decreases accordingly. Type III SRBs usually occur in groups of 3 to 10, lasting less than 60 seconds. For ease of distinction, this study will detect Type III SRBs that appear in groups as a separate category called Type IIIs SRBs. Type IV \citep{Mann1989Interpretation} SRBs are characterized by continuous radiation, which manifests as a steady enhancement of background levels over a wide frequency band. Sometimes, they exhibit extreme continuous radiation covering more than 300 MHz. Type V \citep{Neylan1959An} SRBs are always associated with Type III SRBs and last for several tens of seconds. They appear like a low-frequency "flag" attached to Type III SRBs, with quasi-continuous emission and a narrower bandwidth than Type III SRBs. The duration of Type V SRBs is usually less than one minute, and they are excited by electron beams scattered at high altitudes \citep{monstein2011catalog}.

The e-CALLISTO \citep{Benz2009A} is a global solar observation network designed to monitor solar activity using the radio spectrum. It is primarily used to observe SRBs and associated coronal mass ejections (CMEs). The e-CALLISTO website supports research in solar physics, allowing researchers to access public data to analyze patterns of solar activity and its effects on Earth. The data for this study comes entirely from the e-CALLISTO website. First, we collected publicly available FIT files of solar radio spectrogram images from the e-CALLISTO website using a crawler script (which is also open-source on Zenodo \citep{wang2025ssacodetr} ) to create a dataset. The dataset contains solar radio spectrogram images of SRBs that occurred from August 2020 to August 2022. Next, we used pyCALLISTO \citep{pawase2020pycallisto} to parse the collected FIT files into JPG images using the default settings. Finally, we manually annotated the JPG images using the LabelImg \citep{Russell2008LabelMe:} tool, strictly following the event record files of solar radio burst provided on the website. The annotations include information on the categories and locations of the SRBs. As shown in Table \ref{tab:tab1}, the dataset was randomly divided into a training set and a validation set in a ratio of 7:3. It is noteworthy that multiple SRB instances may exist within the same spectrogram image. Therefore, the total number of SRB instances exceeds the number of spectrogram images. The total number of annotated SRB instances reached 10,822.

\begin{table*}[h!]
    \caption{Solar Radio Burst Spectrogram Dataset
    \label{tab:tab1}}
      \centering
    \begin{tabular}{lccccccc}
            \hline
            \hline
        \textbf{} & \textbf{Number of images} & \textbf{Number of Instances} & \textbf{{\footnotesize II}} & \textbf{{\footnotesize III}} & \textbf{{\footnotesize III}s} & \textbf{{\footnotesize IV}} & \textbf{{\footnotesize V}} \\
            \hline
        Train set & 6,126                     & 7,584                        & 338                         & 5,736                        & 1,296                         & 47                          & 167                        \\
            \hline
        Val set   & 2,626                     & 3,238                        & 126                         & 2,474                        & 561                           & 18                          & 70                         \\
          \hline
    \end{tabular}
\end{table*}

On the e-CALLISTO website where we collected our data, the duration of a spectrogram image is segmented into 15-minute intervals. Due to the prolonged duration of Type I SRBs, the same Type I SRBs can appear in multiple consecutive spectrogram images. Consequently, it is not feasible to label Type I SRBs in the sample data. Thus, this study does not include the detection of Type I SRBs. This choice is a limitation of this study. In the future, we are considering stitching these spectrogram images together to annotate type I solar radio bursts, thereby creating a more comprehensive dataset of SRBs.

\section{Methods}

\begin{figure}[htb]
    \centering
    \includegraphics[width=1.0\linewidth]{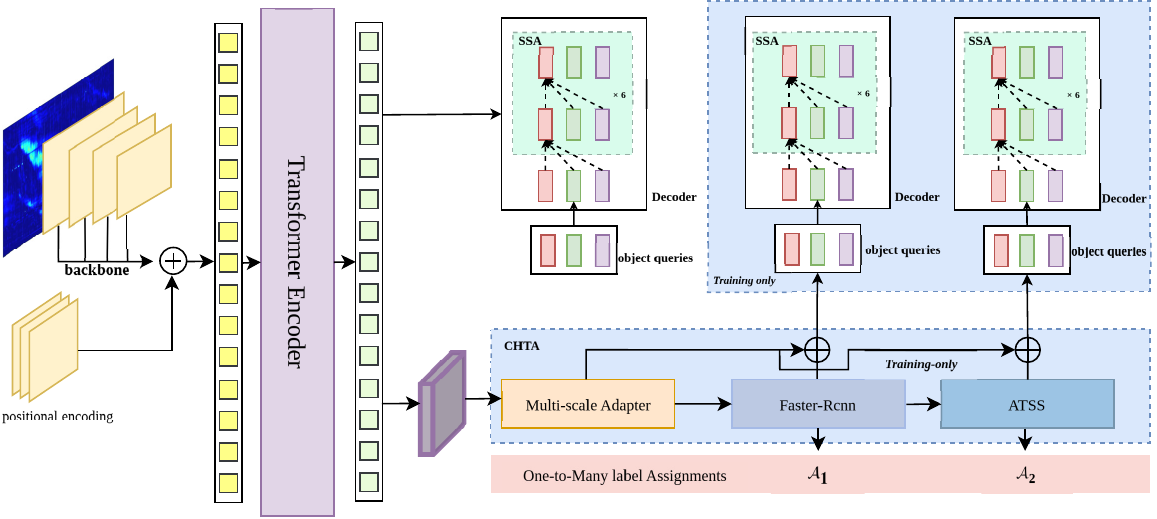}
    \caption{The model's overall structure comprises the backbone, the encoder-decoder structure, and collaborative hybrid auxiliary training (CHAT). The Transformer encoder uses multi-scale attention for encoding, while the decoder uses Scale Sensitive Attention (SSA) for decoding. Collaborative auxiliary training updates the decoder parameters only during training. It does not participate in the forward propagation.} \label{fig:fig2}
\end{figure}

The complexity of SRB scales and strong noise interference are challenges for SRB detection. Traditional detection models are not effective in addressing these issues. Therefore, this paper improves deformable DETR by introducing scale-sensitive attention (SSA), a more suitable attention module. Collaborative hybrid auxiliary training is also introduced to construct a high-performance SRBs detection model, as illustrated in Figure \ref{fig:fig2}. The model uses the backbone network ResNet50 to extract multi-scale features from SRB spectrograms. The encoder-decoder structure predicts the regression of image bounding boxes and classification scores. The SSA module replaces the original attention module in the decoder part to better handle the complex scale variations and noise interference in SRBs. The collaborative hybrid auxiliary training module mitigates performance degradation caused by the positive-negative sample imbalance in the decoder.

Figure \ref{fig:fig2} presents an overview of the model. Solar radio burst spectrogram images are processed through a backbone network to extract multi-scale features. The images are divided into patches after combining these features with positional encoding. These segmented patches are input into the transformer encoder for attention calculations, resulting in intermediate feature memory. During training, the memory is restored to multi-scale features. These features then enter the Faster R-CNN and ATSS auxiliary heads to select positive and negative sample object queries. The selected queries are passed into the transformer decoder for cross-attention calculations, producing predicted bounding boxes and classifications. The auxiliary head is not used during the prediction process, allowing for direct retrieval of the prediction results.

Our proposed model comprises a backbone, an encoder-decoder structure, and collaborative hybrid auxiliary training in Fig \ref{fig:fig2}. The Transformer encoder uses multi-scale attention for encoding, while the decoder uses scale-sensitive attention (SSA) for decoding. Collaborative auxiliary training updates the decoder parameters only during training. It does not participate in forward propagation.

Despite the significant success of DETR in computer vision, it still has drawbacks. These include a high computational load and being unfriendly to small objects. To address these issues, Deformable DETR introduces several improvements over DETR. Enhancements include the introduction of a deformable attention mechanism. In computer vision tasks, the DETR-base model divides an image into multiple patches, with each small patch referred to as a token representing that region's image features. Attention calculation refers to the process of performing vector operations on the input tensor within the attention mechanism. Deformable attention mechanism allows tokens to interact with only a subset of other tokens during the attention calculation, thereby significantly reducing computational load. Deformable DETR also integrates multi-scale feature extraction, effectively capturing feature information of objects at different scales. Additionally, with the global self-attention mechanism of the Transformer, Deformable DETR can fully utilize global contextual information, which is beneficial for modeling the relationships between objects and improving the overall performance of object detection. Extensive experiments on multiple benchmark datasets demonstrate that Deformable DETR offers several advantages. These include flexibility, multi-scale feature extraction, utilization of global contextual information, and end-to-end training in object detection tasks.

\begin{figure}[htb]
    \centering
    \includegraphics[width=0.8\linewidth]{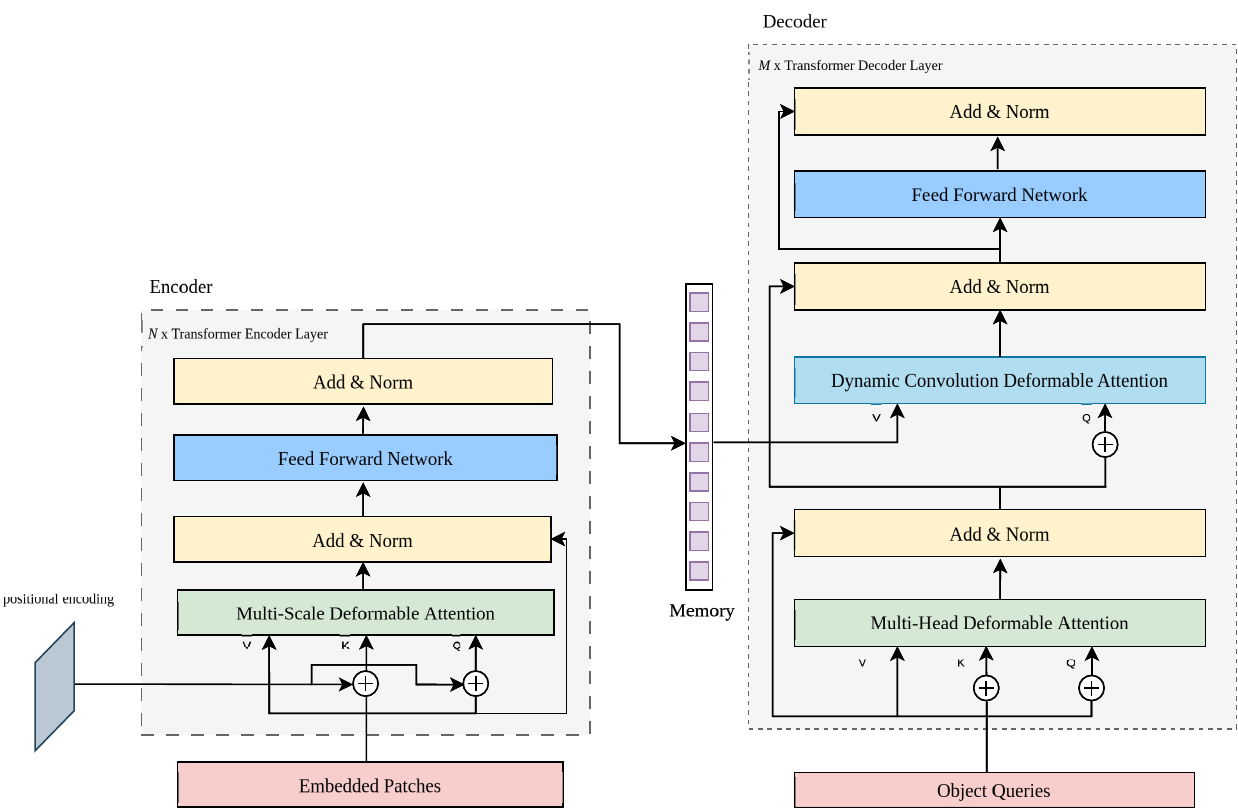}
    \caption{Structure of Deformable DETR. Embedded Patches are derived from feature segmentation obtained from the backbone network. These features are processed through N layers (N=6) of Transformer Encoder Layers to produce intermediate features referred to as Memory. The Memory undergoes M rounds (M=6) of cross-attention with Object Queries to generate the model's predictions.} \label{fig:fig3}
\end{figure}

Deformable DETR adopts an end-to-end training approach. This simplifies the model design and training process, enhancing the model's scalability and generalization. It consists of two main components: the Transformer Encoder Layer and the Transformer Decoder Layer. Figure \ref{fig:fig3} illustrates the model structure of Deformable DETR. On the left side of the diagram is the Encoder, which comprises N (N=6) layers of Transformer Encoder Layers. The input to the Encoder consists of multi-scale image features extracted by the backbone network, referred to as Embedded Patches. In the center of the diagram is the intermediate feature Memory, which is the output from the Encoder. On the right side of the diagram is the Decoder structure, containing M (M=6) layers of Transformer Decoder Layers. After the cross-attention calculation between the Memory and the Object Queries (query vectors) in the Decoder, the final output tensor of the model, representing the predicted results, is obtained.

The Transform Encoder Layer can effectively encode the features of input embedded patches into intermediate features represented as Memory. It mainly consists of multi-scale deformable attention, feed-forward neural networks, layer normalization, and residual connections. Multi-scale deformable attention computes attention at different scales on the input data to capture feature information across various scales. It dynamically adjusts the computation locations of attention weights, enabling the model to better adapt to different scales and shapes of targets. The feed-forward neural network performs nonlinear transformations and mappings on the features at each position, enhancing the model's ability to represent and extract features from the input data. Normalization helps reduce internal covariate shifts, accelerating model convergence and improving generalization. Residual connections prevent the vanishing gradient problem in deep network training.

The Transformer Decoder Layer decodes the encoded features and generates the object detection results. The Decoder Layer uses Object Queries for boundary regression and classification prediction. Object Queries help the model learn feature representations related to object categories. By embedding object categories into Object Queries, the model can interact with the Memory through these vectors, effectively locating objects and generating accurate object detection results.

This paper introduces improvements to Deformable DETR to better adapt to the characteristics of the solar radio burst dataset. The following sections, 5.1 to 5.3, detail the improvements, including the SSA module, the collaborative hybrid auxiliary training module, and the loss function used in this model.

\subsection{Scale Sensitive Attention (SSA)}

The deformable attention mechanism provides better flexibility and modeling capabilities. However, it relies on large-scale training data, which may not offer much advantage for small datasets. Additionally, the excessive flexibility might lead to over-fitting on the training set.

\begin{figure}[htb]
    \centering
    \includegraphics[width=0.8\linewidth]{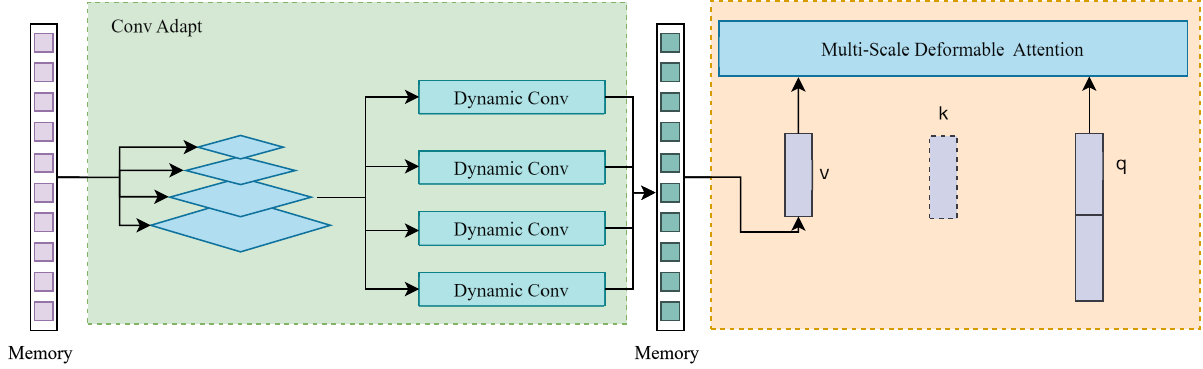}
    \caption{Scale Sensitive Attention Module. Memory refers to the intermediate image features output by the model. After being restored to multi-scale features, these are fed into different dynamic convolutions.} \label{fig:fig4}
\end{figure}

For this application, this paper introduces the SSA module, as shown in Figure \ref{fig:fig4}. Before the multi-scale deformable attention, a convolution adapter is introduced to perform multi-scale dynamic convolution on the value. The introduced convolution module can constrain the model's receptive field and promote the establishment of connections within a narrower range. This enhances the model's ability to recognize the continuity of SRBs while limiting the model's boundary expansion, thereby more precisely adapting to the target. The experiments in this paper validate that SSA can effectively improve the model's performance.

In Figure \ref{fig:fig4}, SSA comprises two main components: the convolution adapter and multi-scale deformable attention.

\begin{figure}[htb]
    \centering
    \includegraphics[width=0.7\linewidth]{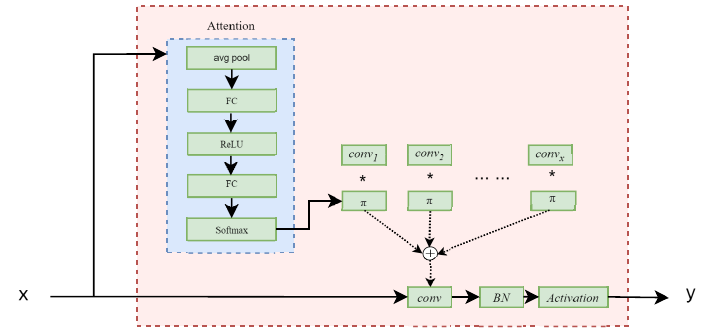}
    \caption{Dynamic Convolution Module. The tensor \( X \) represents the features output by the previous layer of the model. After branching, it is processed separately through Attention and Convolution. The resulting outputs are then combined and passed through an activation function to obtain tensor \( Y \). At this stage, the shape of the vector remains unchanged.} \label{fig:fig5}
\end{figure}

\begin{figure}[htb]
    \centering
    \includegraphics[width=0.6\linewidth]{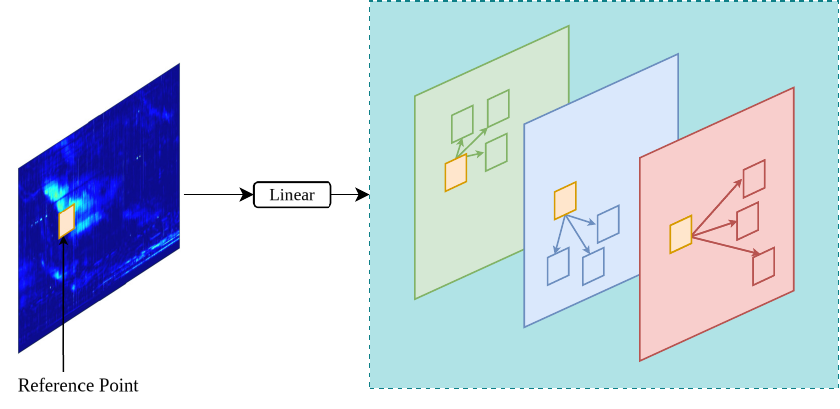}
    \caption{Illustration of Multi-scale Deformable Attention. The reference point represents a token that, after obtaining multi-scale features, no longer interacts with all other tokens. Instead, it dynamically engages in computations with specific other tokens.} \label{fig:fig6}
\end{figure}

The convolution adapter performs multi-scale decomposition on the memory and then feeds it into different dynamic convolutions, with the core component being dynamic convolution, as shown in Figure \ref{fig:fig5}. Multi-scale deformable attention combines multi-scale features and the deformable attention mechanism, as shown in Figure \ref{fig:fig6}. It effectively captures feature information at different scales and enhances model performance.

In Figure \ref{fig:fig5}, dynamic convolution is an adaptive convolution operation. It can dynamically adjust the size and shape of the convolutional kernels to accommodate the features of different input data. Traditional convolution operations use fixed-size kernels for feature extraction, which may not capture features of varying sizes and shapes in the input data. On the other hand, dynamic convolution learns adjustable kernel parameters that dynamically allow kernels to change their size and shape with each convolution operation. This enhances the model's performance and generalization capability when handling complex data and tasks.

In Figure \ref{fig:fig6}, richer and more comprehensive feature information can be captured at different scales by introducing multi-scale features. This helps enhance the model's ability to recognize and locate targets of varying scales. Additionally, the deformable attention mechanism can dynamically adjust attention weights based on the features of the input data. This enables the model to focus more on important feature regions, thereby improving its performance in complex scenarios.

\subsection{Collaborative Hybrid Assignments Training (CHAT)}

In the DETR-based model, the number of queries represents the number of predicted bounding boxes output by the decoder. The optimal value, determined through multiple experiments in the Deformable DETR paper, is 300. Randomly assigning queries performs well on the COCO dataset, which averages 7.7 targets per image. However, it performs poorly on the SRB dataset, which averages only 1 to 2 targets per image, resulting in many predicted bounding boxes for negative samples. Therefore, this paper introduces Collaborative Hybrid Assignments Training into (CHAT) the model.

\begin{figure}[htb]
    \centering
    \includegraphics[width=0.8\linewidth]{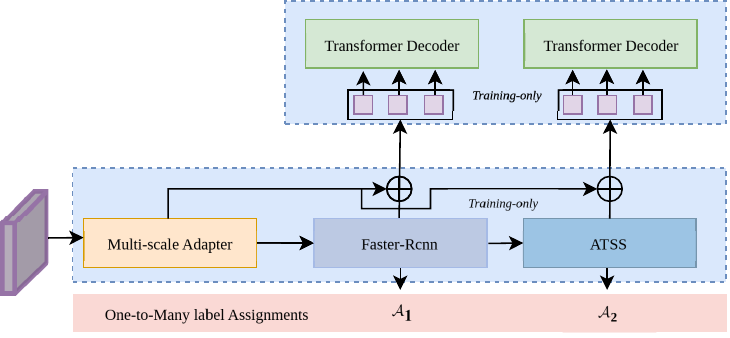}
    \caption{Collaborative Hybrid Assignments Training. The input to the multi-scale adapter consists of the restored multi-scale features. These multi-scale features are processed through auxiliary heads, namely Faster-RCNN and ATSS, to obtain the selected object queries. The object queries are then fed into the Decoder for further update and training.} \label{fig:fig7}
\end{figure}
Collaborative Hybrid Assignments Training is a training method for DETR-based detectors. In DETR, which employs one-to-one set matching, having too few queries assigned as positive samples leads to sparse supervision for the encoder's output. This significantly hampers the encoder's ability to learn discriminative features, as illustrated in Figure \ref{fig:fig7}. Collaborative Hybrid Assignments Training is only used during training and does not participate in forward propagation. It alleviates the issue of sparse supervision by using traditional neural network outputs with one-to-many label assignments as the decoder's object queries. Collaborative Hybrid Assignments Training effectively adapts to having few targets in SRBs detection.

\begin{figure}[htb]
    \centering
    \includegraphics[height=4in]{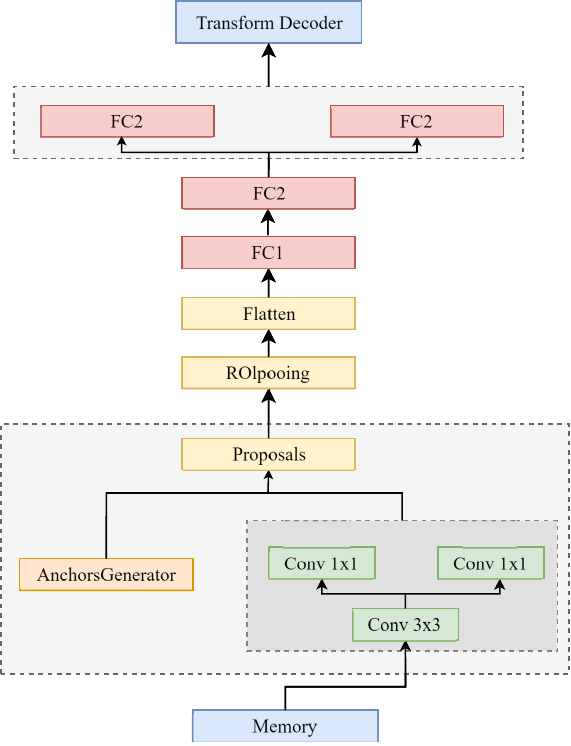}
    \caption{Faster R-CNN auxiliary head, where Cross-Entropy (CE) loss is used as the classification loss and Generalized Intersection over Union (GIoU) loss is used as the regression loss. Predictions with a GIoU less than 0.5 are considered negative samples, while those with a GIoU greater than or equal to 0.5 are considered positive samples.} \label{fig:fig8}
\end{figure}

In Figure \ref{fig:fig7}, two one-to-many auxiliary heads are used: Faster R-CNN and ATSS. These will be introduced in the following sections.

The Faster R-CNN auxiliary head is an improved version of R-CNN and Fast R-CNN, as shown in Figure \ref{fig:fig8}. The Faster R-CNN auxiliary head introduces the Region Proposal Network (RPN), which allows the generation of candidate regions to share the full convolutional network (FCN) with the detection network, significantly enhancing detection speed and performance.

In Figure \ref{fig:fig7}, the adaptive training sample selection (ATSS) auxiliary head is an adaptive mechanism for selecting training samples. It aims to address the issue of imbalanced positive and negative samples in object detection by dynamically adjusting the criteria for defining positive and negative samples to optimize the model's training process. The computation process can be represented as follows:
\begin{equation}
    {{t}_{g}}=Mean(IoU({{C}_{g}},g))+Std(IoU({{C}_{g}},g))
    \label{equ:ATSS}
\end{equation}
where, ${{t}_{g}}$  represents the threshold for classifying positive and negative samples,  $Mean$ represents the mean calculation, and $Std$ represents the variance calculation. $IoU$ (Intersection over Union) indicates the overlap between the predicted bounding box and the ground truth bounding box, where ${{C}_{g}}$ denotes the predicted box, and $g$ denotes the ground truth box.

Figure \ref{fig:fig8} illustrates the structure of Faster R-CNN. The intermediate feature "Memory" is processed through convolution and then stacked with manually designed anchors before entering the pooling layer. The resulting output is then flattened (the flatten operation transforms high-dimensional input data into a one-dimensional vector) and passed through FC1 (the first fully connected layer in the neural network) and FC2 (the second fully connected layer) to produce the final output.

\subsection{Loss Function}
The overall loss function of the model consists of three components: the decoder loss, the auxiliary head decoder loss, and the auxiliary head loss \citep{zong2023detrs}, which can be expressed as:
\begin{equation}
    {{\mathcal{L}}^{global\text{ }\!\!~\!\!\text{ }}}=\underset{l=1}{\overset{L}{\mathop \sum }}\,\left( \tilde{\mathcal{L}}_{l}^{dec\text{ }\!\!~\!\!\text{ }}+{{\lambda }_{1}}\underset{i=1}{\overset{K}{\mathop \sum }}\,\mathcal{L}_{i,l}^{dec}+{{\lambda }_{2}}{{\mathcal{L}}^{enc}} \right)
    \label{equ:total loss}
\end{equation}
where, $l$ denotes the index of the decoder layer $l\in (0,L)$, while${{\lambda }_{1}}$and${{\lambda }_{2}}$ are balancing coefficients. $i$ represents the index of the auxiliary head $i\in (0,K)$ (set to 2 in this experiment). $\mathcal{L}_{i,l}^{dec\text{ }}$ denotes the loss of the $l$-th auxiliary head in the $i$-th decoder layer, which can be expressed as:
\begin{equation}
    \mathcal{L}_{i,l}^{dec}=\tilde{\mathcal{L}}\left( {{{\tilde{P}}}_{i,l}},P_{i}^{\left\{ pos \right\}} \right)
    \label{equ:loss1}
\end{equation}
where, ${{\tilde{P}}_{i,l}}$ represents the output prediction of the $l$-th auxiliary head in the $i$-th decoder layer. $\mathcal{L}^{enc\text{ }}$represents the loss of the $K$ auxiliary heads, which can be expressed as:
\begin{equation}
    {{\mathcal{L}}^{enc}}=\underset{i=1}{\overset{k}{\mathop{\sum }}}\,{{\text{}}_{i}}\left( \hat{P}_{i}^{\left\{ pos \right\}},P_{i}^{\left\{ pos \right\}} \right)+{{\mathcal{L}}_{i}}\left( \hat{P}_{i}^{\left\{ neg \right\}},P_{i}^{\left\{ neg \right\}} \right)
    \label{equ:loss2}
\end{equation}
where, $P_{i}^{\{pos\}}$ denotes the supervision target for positive samples of the $i$-th auxiliary head, including both the category and regression offset. $\hat{P}_{i}^{\left\{ pos \right\}}$ represents the prediction of the $i$-th auxiliary head for the positive samples. $\tilde{\mathcal{L}}_{l}^{dec}$ stands for the loss in the original one-to-one set matching \citep{carion2020end} , which can be expressed as:
\begin{equation}
    \tilde{\mathcal{L}}_{l}^{dec\text{ }\!\!~\!\!\text{ }}\left( y,\hat{y} \right)=\underset{i=1}{\overset{N}{\mathop{\sum }}}\,\left[ -\log {{{\hat{p}}}_{\hat{\sigma }(i),l}}\left( {{c}_{i}} \right)+{{\mathcal{L}}_{box\ }}\left( {{b}_{i}},{{{\hat{b}}}_{\hat{\sigma },l}}(i) \right) \right]
    \label{equ:loss3}
\end{equation}
where, $N$ is defined as the number of predictions made by the decoder. The prediction for each real set element $i$ can be considered as ${{y}_{i}}=\left( {{c}_{i}},{{b}_{i}} \right)$, where ${{c}_{i}}$ is the target class label (the set is $\varnothing $).${{b}_{i}}\in {{[0,1]}^{4}}$ represents the ground truth bounding box, which includes the coordinates of the center point and the width and height of the box. $-\log {{\hat{p}}_{\hat{\sigma }(i),l}}\left( {{c}_{i}} \right)$ is defined as the classification loss, and ${{\mathcal{L}}_{box\ }}\left( {{b}_{i}},{{{\hat{b}}}_{\hat{\sigma },l}}(i) \right)$ is defined as the localization loss, which can be expressed as:
\begin{equation}
    {{\mathcal{L}}_{box\text{ }\!\!~\!\!\text{ }}}\left( {{b}_{i}},{{{\hat{b}}}_{\hat{\sigma },l}}\left( i \right) \right)={{\lambda }_{iou\text{ }\!\!~\!\!\text{ }}}{{\mathcal{L}}_{iou\text{ }\!\!~\!\!\text{ }}}\left( {{b}_{i}},{{{\hat{b}}}_{\sigma (i),l}} \right)+{{\lambda }_{L1}}\parallel {{b}_{i}}-{{\hat{b}}_{\sigma (i),l}}{{\parallel }_{\text{ }1}}
    \label{equ:loss4}
\end{equation}
where, $\parallel {{b}_{i}}-{{\hat{b}}_{\sigma (i),l}}\parallel $ represents the loss calculated as the absolute difference between the predicted bounding box's center coordinates, height, and width, and the corresponding values of the ground truth bounding box. ${{\mathcal{L}}_{iou\text{ }\!\!~\!\!\text{ }}}$ is the localization loss based on the Intersection over Union (IoU) between the predicted bounding box and the ground truth bounding box.

\begin{figure}[htb]
    \centering
    \includegraphics[width=0.2\linewidth]{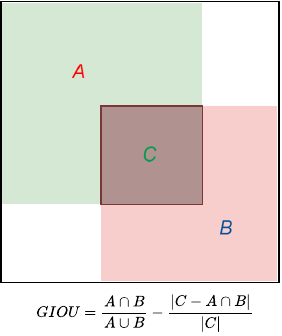}
    \caption{Example of Generalized Intersection over Union (GIoU).}\label{fig:fig9}
\end{figure}
The IoU loss aims to directly optimize the IoU value between the predicted bounding box and the ground truth bounding box, thereby improving the model's accuracy in bounding box localization. Figure \ref{fig:fig9} illustrates the localization loss function, Generalized Intersection over Union (GIoU), used in this paper.

\section{Experimental Results and Analysis}

\subsection{Experimental Software and Hardware Environment and Setup}

All experiments in this paper were conducted on hardware consisting of an Intel i5-13400F CPU, an NVIDIA GeForce RTX 4090 GPU, and 32GB of memory. The software environment was Windows 11, with PyTorch \citep{paszke2019pytorch} 1.11.0 and MMDetection \citep{chen2019mmdetection} v2.25 installed.

The learning rate is a hyperparameter that controls the step size at which the model updates its weights during training. Weight decay is a regularization technique used to prevent model overfitting. It does this by adding a penalty term to the loss function, thereby constraining the model's complexity. The initial learning rate was set to 2e-05, and the optimizer used was AdamW with a weight decay of 2e-05. For the learning rate decay strategy, this paper employed the cosine annealing learning rate method, setting the training epochs to 24 to gradually reduce the learning rate and improve model stability.

\subsection{Evaluation Metrics}
The performance evaluation metrics used in this paper's experiments include AP50, Recall, FLOPs, and Params.

The detection accuracy metric used is AP50. AP stands for Average Precision, which refers to the average precision across all classes in multi-class prediction scenarios. AP50 refers to the area under the precision-recall curve when the Intersection over Union (IoU) threshold is set to 0.5. This metric comprehensively reflects the combined performance of precision and recall, thereby thoroughly evaluating the model's performance in recognizing specific categories. The calculation methods for precision and recall are as follows:

\begin{equation}
    precision=\frac{TP}{TP+FP}
    \label{equ:metrics1}
\end{equation}
\begin{equation}
    recall=\frac{TP}{TP+FN}
    \label{equ:metrics2}
\end{equation}
where, TP represents the number of detected boxes with $IoU > 0.5$, FP represents the number of detected boxes with $IoU \le 0.5$ and the number of redundant boxes, and FN represents the number of missed target boxes.

Floating Point Operations (FLOPs) are used to measure the complexity of an algorithm and are often used as an indirect measure of the speed of neural network models. Parameters (Params) refer to the total number of trainable parameters in the model, which are used to assess the model's spatial complexity.

\subsection{Comparative experiments with other models}

The proposed model was compared with several classical models and models previously proposed by researchers. These experiments were conducted on the dataset we constructed (details in Section 4 and Table \ref{tab:tab1}). The models compared include YOLOv3 with MobileNetv2 as the backbone network \citep{oguine2022yolo}. Another model is YOLOv5s \citep{Wang2022A} with K-means anchors. Furthermore, MV2-SSDLite uses generalized focal loss (GFL) \citep{li2020generalized} as the loss function, combined with Adaptive Training Sample Selection (ATSS). Then, SSDLite \citep{hoops2024universal} with SwinTransformer Tiny as the backbone network is also compared. Additionally, comparisons were made with methods using SSDLite as the detection head and Swin Transformer Tiny as the feature extractor, as well as methods using SSDLite as the detection head and MobileVit as the feature extractor. Finally, the proposed model was compared with the DETR-base model, including the original DETR model and the recently proposed Dino-DETR \citep{zhang2022dino}. The results of the comparative experiments are shown in Table \ref{tab:tab2}.

\begin{table}\centering
    \caption{ Comparative Experimental Results with Other Models. AP@50 refers to the average precision when the IoU threshold is set to 0.5, with higher values indicating greater model accuracy. Recall measures the model's ability to correctly identify all positive samples. A higher recall value signifies stronger detection capability for positive samples. FLOPs evaluate the number of floating point operations required during model computation, with a lower value indicating a lighter model. Params refer to the total number of learnable parameters in the model. A larger number implies a stronger learning capacity but may also lead to a higher risk of overfitting.} \label{tab:tab2}
    \begin{tabular}{lcccccccccc}
        \hline
        \hline
        \multirow{2}{*}{\textbf{Model}} & \multicolumn{6}{c}{\textbf{{mAP50@50}}} & \multirow{2}{*}{\textbf{Recall}}                           & \multirow{2}{*}{\textbf{FLOPs}}                            & \multirow{2}{*}{\textbf{Params}}                                                                                                                                                                                                              \\
        \cline{2-7}                     & \multicolumn{1}{c}{Average}             & \multicolumn{1}{c}{\uppercase\expandafter{\romannumeral2}} & \multicolumn{1}{c}{\uppercase\expandafter{\romannumeral3}} & \multicolumn{1}{c}{\uppercase\expandafter{\romannumeral3}s} & \multicolumn{1}{c}{\uppercase\expandafter{\romannumeral4}} & \multicolumn{1}{c}{\uppercase\expandafter{\romannumeral5}}                                                         \\

        \hline
        YOLOv3                          & 0.738                                   & 0.774                                                      & 0.827                                                      & 0.814                                                       & 0.517                                                      & 0.759                                                      & 89.5\%          & 1.69G          & 3.74M              \\

        GFL                             & 0.742                                   & 0.806                                                      & 0.798                                                      & 0.819                                                       & 0.530                                                      & 0.756                                                      & 96.6\%          & 20.78G         & 32.04M             \\

        MV2-SSDLite                     & 0.75                                    & 0.811                                                      & 0.787                                                      & 0.777                                                       & 0.652                                                      & 0.721                                                      & 88.8\%          & \textbf{0.71G} & \textbf{3.05M}     \\

        YOLOV5                          & 0.752                                   & 0.753                                                      & 0.867                                                      & 0.809                                                       & 0.566                                                      & 0.766                                                      & 69.5\%          & 7.943G         & 7.033M             \\

        SwinT-SSDLite                   & 0.764                                   & 0.810                                                      & 0.842                                                      & 0.781                                                       & 0.714                                                      & 0.689                                                      & 90.8\%          & 9.82G          & 28.79M             \\

        MobileVit-SSDLite               & 0.782                                   & 0.827                                                      & 0.827                                                      & 0.803                                                       & \textbf{0.725}                                             & 0.728                                                      & 92\%            & 2.9G           & 5.32M              \\

        DETR                            & 0.745                                   & 0.732                                                      & 0.872                                                      & 0.776                                                       & 0.632                                                      & 0.714                                                      & 97.3\%          & 225G           & 41M                \\

        Dino-DETR                       & 0.741                                   & 0.777                                                      & 0.88                                                       & 0.805                                                       & 0.591                                                      & 0.684                                                      & 98\%            & 86G            & 47M                \\

        DETR4SBRs(Ours)                            & \textbf{0.835}                          & \textbf{0.873}                                             & \textbf{0.924}                                             & \textbf{0.842}                                              & 0.714                                                      & \textbf{0.821}                                             & \textbf{99.4\%} & 113G           & 73.67M(57M$\star$) \\
         \hline
    \end{tabular}
  \end{table}

As shown in Table \ref{tab:tab2}, DETR4SBRs proposed in this study achieved higher average detection accuracy. It is slightly weaker than MobileViT-SSDLite in detecting Type IV bursts, but it outperforms other models in detecting other burst types.

Since automated SRB detection is primarily used for initial screening to filter out large amounts of non-burst data, a few burst events can be manually confirmed in the final step. Therefore, in this application scenario, it is crucial to minimize missed detections of SRBs, making the recall rate particularly important. The method proposed in this study achieved a recall rate of 99.4$\%$, outperforming all other models. Regarding model parameters, the DETR-based model relies on the Transformer architecture, an advanced deep-learning model framework. It demonstrates powerful performance by incorporating complex components such as multi-head attention mechanisms and positional encoding. However, these components also contribute to the complexity of the Transformer model, requiring more parameters and computational resources for training and inference, which is a significant disadvantage compared to lightweight models. Compared horizontally with the DETR-base model, DETR4SBRs proposed in this study shows better performance regarding floating-point operations per second (FLOPs).

\subsection{Analysis of Model Improvements}
\subsubsection{Model Improvement Effectiveness Analysis}

\begin{table}[htb]
    \centering
    \caption{Ablation Studies. CHAT stands for Collaborative Hybrid Assignments Training. SSA stands for Scale Sensitive Attention. AP@50 refers to the average precision when the IoU threshold is set to 0.5, with higher values indicating greater model accuracy. Recall measures the model’s ability to correctly identify all positive samples. A higher recall value signifies stronger detection capability for positive samples. } \label{tab:tab3}
    \begin{tabular}{lcccc}
      
        \hline
        \hline

        \textbf{CHAT}     & \textbf{SSA}      & \textbf{GIoULoss} & \textbf{mAP@50} & \textbf{Recall} \\
         \hline
                          &                   &                   & 0.65            & 97\%            \\

        \text{\checkmark} &                   &                   & 0.795           & \textbf{99.5\%} \\
                          & \text{\checkmark} &                   & 0.726           & 96.3\%          \\
                          &                   & \text{\checkmark} & 0.629           & 99.1\%          \\
        \text{\checkmark} & \text{\checkmark} &                   & 0.820           & 99.2\%          \\
                          & \text{\checkmark} & \text{\checkmark} & 0.750           & 97.1\%          \\
        \text{\checkmark} &                   & \text{\checkmark} & 0.817           & 98.4\%          \\
        \text{\checkmark} & \text{\checkmark} & \text{\checkmark} & \textbf{0.835}  & 99.4\%          \\
         \hline
    \end{tabular}
\end{table}

We conducted comparative experimental analyses on the proposed improvements. Multiple improved models were constructed sequentially, and the results were compared using the same test data. The performance enhancements brought by the added modules are demonstrated in Table \ref{tab:tab3}. The model's mAP@50 accuracy was relatively low at only 65$\%$ without any improvements. After adding only the Collaborative Hybrid Assignments Training, the mAP@50 increased to 79.5$\%$. With only the SSA module added, the mAP@50 improved to 72.6$\%$. However, we observed that the model's accuracy decreases when only the GIoULoss is added. The model performs better when working in conjunction with other modules. We realize that this may be due to the sensitivity of GIoULoss to noise, and the original model's weak ability to extract image features does not effectively eliminate the noise present in the images. When all improvements were incorporated, the mAP@50 reached its highest at 83.5$\%$. Here, the recall rate including collaborative auxiliary hybrid training reaches a maximum of 99.5\%. We believe that the decrease in recall after adding other modules is due to the stricter criteria applied by the GIoU method to the detection boxes.

\subsubsection{Data Augmentation}
\begin{figure}[htb]
    \centering
    \includegraphics[width=0.7\linewidth]{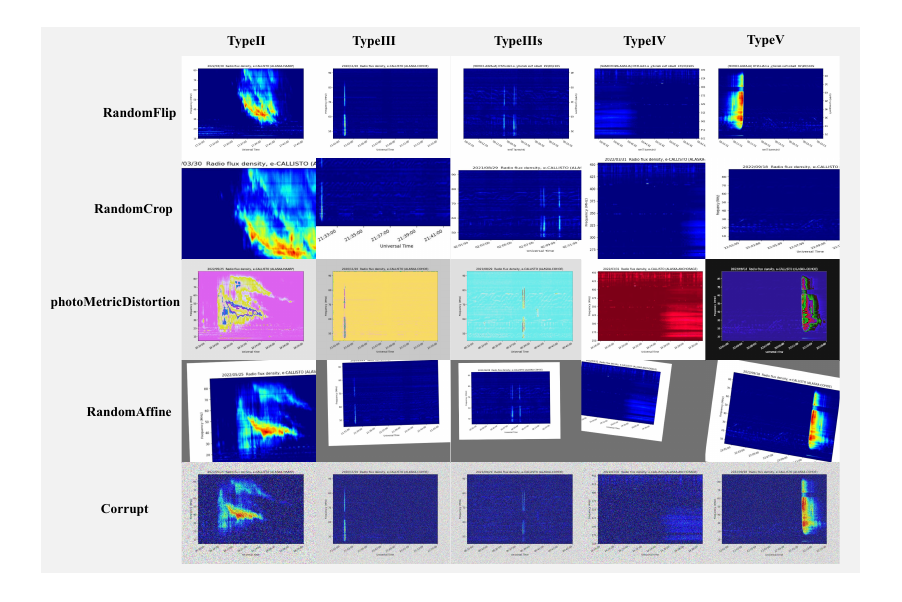}
    \caption{Illustrations of various data augmentations on solar radio burst types II, III, IIIs, IV, and V.} \label{fig:fig10}
\end{figure}

Data augmentation is a technique employed for training machine learning models. It creates new training samples by applying various transformations and processing methods to the existing data. As the number of model parameters increases, when the dataset is small, the model may "memorize" all the images in the training set instead of learning the intrinsic features of the images. This phenomenon is known as overfitting, and data augmentation can effectively mitigate this issue. Additionally, data augmentation can enhance the generalization ability and robustness of the model. For instance, in the third row of Figure \ref{fig:fig14}, we converted the FIT files into grayscale images, and the model still performed well. This might be attributed to the photoMetricDistortion image augmentation method, which helps the model better understand the concept of "background" rather than recognizing a specific shade of blue as the background. This study explores the impact of different types of data augmentation on model performance by applying a series of random transformations to the original images. This can generate new training samples, helping the model learn more robust features, as illustrated in Figure \ref{fig:fig10}.

Figure \ref{fig:fig10} illustrates various data augmentation methods used in this study. These include random flipping, random cropping, photometric distortions, random affine transformations, and image corruptions. These augmented images are used to train the model, reducing the risk of overfitting and enhancing the model's robustness.

\begin{figure}[htb]
    \centering
    \includegraphics[width=0.6\linewidth]{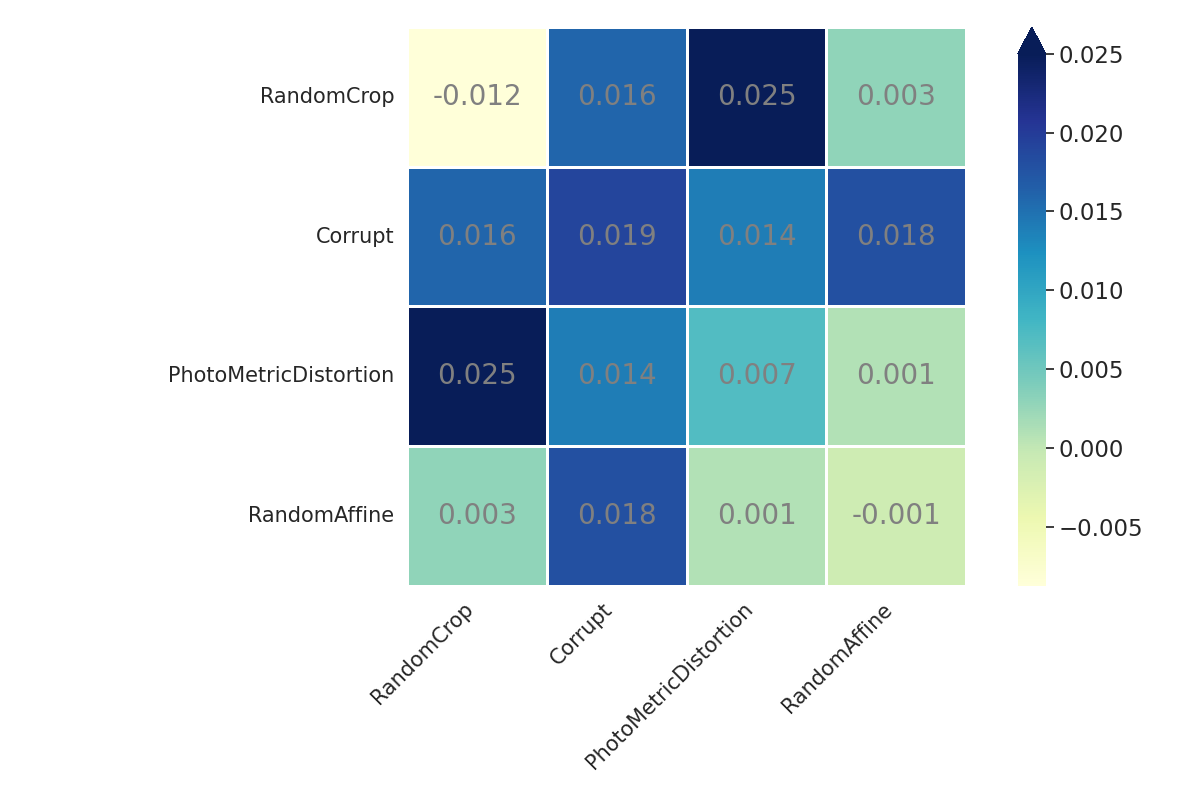}
    \caption{Performance of Image Augmentation. Note: Random flipping is included by default in each experiment. The position of the diagonal indicates that only a single augmentation is applied.} \label{fig:fig11}
\end{figure}

Random flipping involves randomly flipping an image either horizontally or vertically. Random cropping selects a random region from the original image and crops it to create a new image. This method simulates scenarios where parts of the object are occluded by changing the size and aspect ratio of the object. Photometric distortions involve altering the brightness, contrast, saturation, and hue of an image to create a new one. Random affine transformations can simulate geometric errors in instrument observations through linear transformations. For example, rotation can be used to reflect the calibration errors in the antenna polarization direction or the directional distortion of dynamic spectra (such as the inclination of burst stripes) caused by the relative motion between the Sun and Earth. Image corruptions introduce noise and degradation effects (such as blurring, pixelation, and JPEG compression artifacts).

Compared to using all image augmentation techniques, targeted data augmentation has a greater impact on improving mAP@50. This study conducted comparative experiments to identify which data augmentations effectively enhance the model's mAP@50. In Figure \ref{fig:fig11}, the X-axis and Y-axis represent the types of augmentations, and the corresponding values indicate the changes in mAP@50 when using different data augmentations. The combination of random cropping and photometric distortion yields the best results. This experiment utilizes this combination for data augmentation. In contrast, using only random cropping results in a noticeable decrease in accuracy.

\subsubsection{Analysis of the Scale Sensitive Attention Module}
\begin{figure}[htb]
    \centering
      \includegraphics[width=1.0\linewidth]{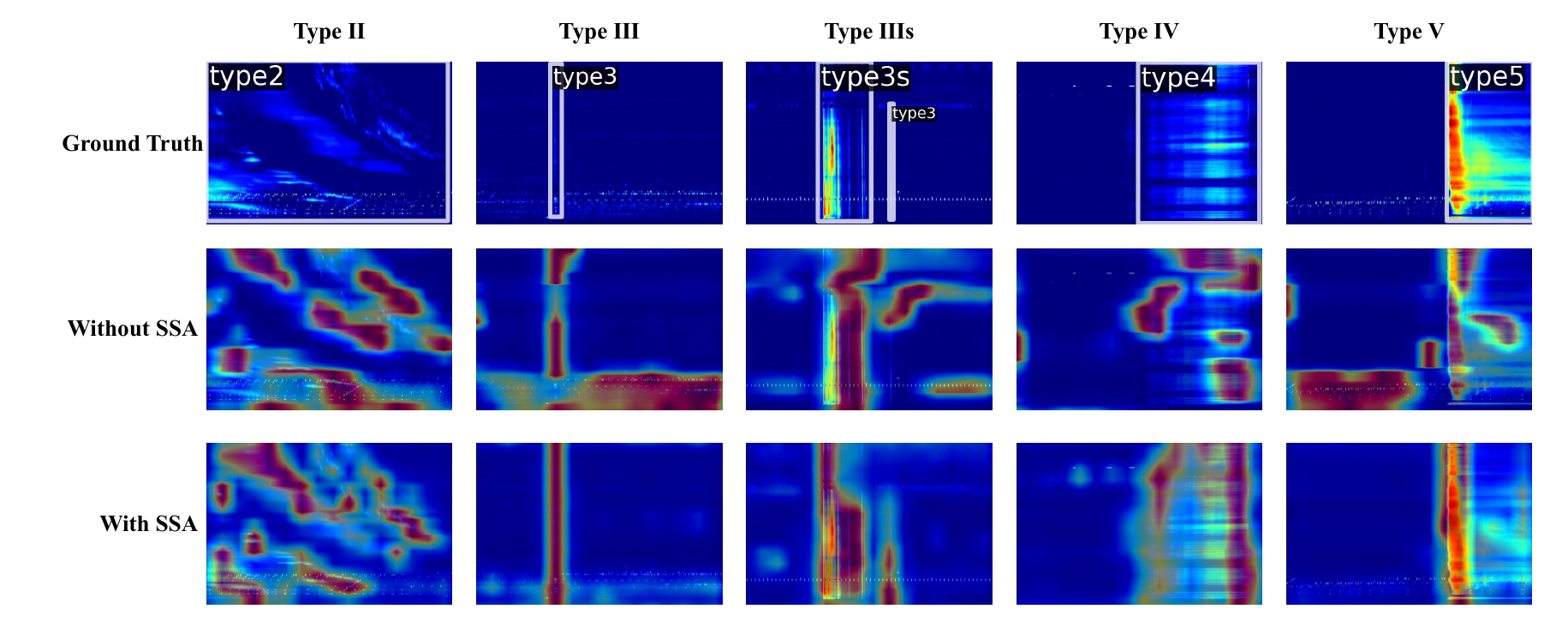}
       \caption{GradCAM Heatmap Comparison of Solar Radio Burst Spectrogram Images. Where, red indicates high attention areas and blue represents low attention areas.} \label{fig:fig12}
\end{figure}

This study uses GradCAM \citep{selvaraju2017grad} for heatmap analysis to compare the attention distribution with and without SSA, as shown in Figure \ref{fig:fig12}. GradCAM (Gradient-weighted Class Activation Mapping) is a visualization technique designed to understand which areas of an image the neural network model focuses on when making decisions. Grad-CAM combines the weights with the feature maps from convolutional layers to create a weighted feature map. This map is then processed through a ReLU activation function to eliminate negative values, and the results are scaled to the original image dimensions to produce the final heatmap. The generated heatmap is overlaid on the original image with color encoding, where red indicates high-attention areas and blue represents low-attention areas. This allows for a clear visual representation of the regions the model prioritizes when making classification decisions.

As shown in Figure \ref{fig:fig12}, incorporating the SSA module enables the model to better focus on the solar radio bursts rather than paying excessive attention to the background. When observing the heatmaps, it is necessary to refer to the ground-truth images in the first row. In the column of type II, it can be seen that the model with the SSA module focuses on areas that are more consistent with the shape of the bursts. In the column of type III, it is more evident that the model with the SSA module mainly focuses on the bursts, while the model without the SSA module pays attention to the noise in the background. This phenomenon is also observed in the columns of type IIIs, type IV, and type V. This indicates that the SSA module can effectively reduce the probability of the appearance of redundant bounding boxes.

\subsubsection{Analysis of the Auxiliary Heads}

This study conducted further ablation experiments on the auxiliary heads. As shown in Table \ref{tab:tab4}, the auxiliary ATSS improves the model's performance.

\begin{table}[htb]
    \centering
    \caption{Ablation Studies on Auxiliary Heads. In this context, Co\_Faster\_Rcnn refers to the auxiliary head of Faster\_Rcnn, while Co\_ATSS denotes the auxiliary head of ATSS. AP@50 refers to the average precision when the IoU threshold is set to 0.5, with higher values indicating greater accuracy of the model. Similarly, a higher value of recall signifies an improved capability of the model to detect positive samples.} \label{tab:tab4}
    \begin{tabular}{lcccc}
        \hline
        \hline

        \textbf{Co\_Faster\_Rcnn} & \textbf{Co\_ATSS} & \textbf{mAP@50} & \textbf{Average recall} \\
        \hline
        \text{\checkmark}         &                   & 0.773           & \textbf{99.8\%}         \\
                                  & \text{\checkmark} & 0.791           & 98.6\%                  \\
        \text{\checkmark}         & \text{\checkmark} & \textbf{0.795}  & 99.5\%                  \\

      \hline
    \end{tabular}
\end{table}

\subsection{Visualization of Detection Results}

\begin{figure}[htb]
    \centering
    \includegraphics[width=0.8\linewidth]{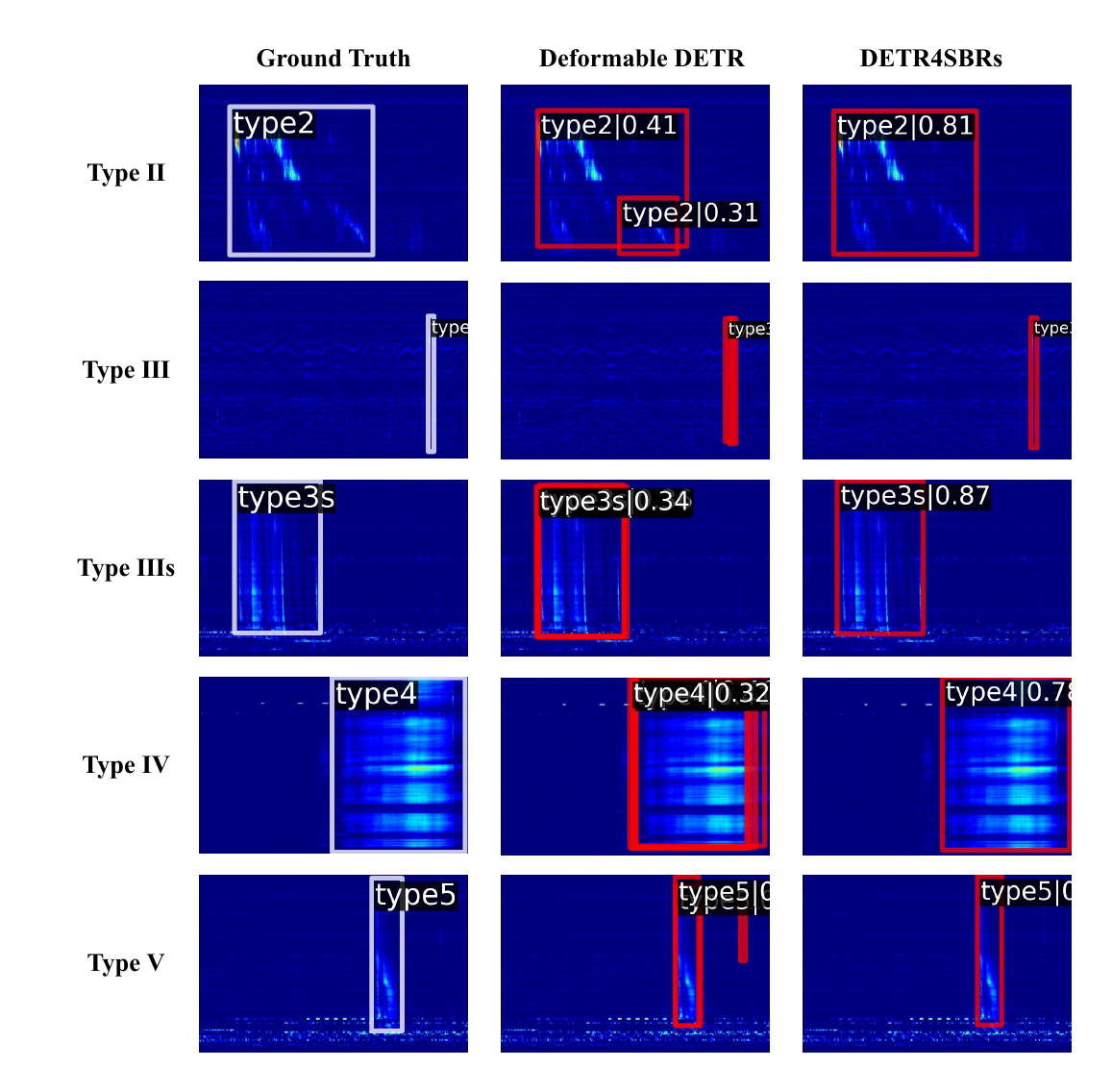}
    \caption{Detection Results of Solar Radio Burst Types II, III, IIIs, IV, and V. It can be observed that DETR4SBRs surpasses the original method in recognizing multiple categories. DETR4SBRs demonstrates greater sensitivity to the continuity of eruptions and exhibits more confidence in its bounding box predictions.}
    \label{fig:fig13}
\end{figure}

Figure \ref{fig:fig13} visualizes the detection results for different types of SRBs in this study, including Types II, III, IIIs, IV, and V.

In Figure 13, the blue boxes in the leftmost column represent the annotated ground truth boxes. The middle column shows the detection results using Deformable DETR. The right column displays the detection results using the improved method proposed in this paper. The red boxes in the middle and right columns indicate the detection results, including the detected positions, SRB types, and confidence information of SRBs. The proposed method shows higher confidence and fewer false detections in detecting various types of SRBs. Due to the longer duration of Type IV SRBs, the Deformable DETR model generates redundant boxes, whereas DETR4SBRs proposed in this paper does not produce any redundant boxes. Additionally, DETR4SBRs achieves more precise localization regression and higher confidence for some finer SRB categories.

\subsection{Analysis of Model Robustness}

\begin{figure}[htb]
    \centering
    \includegraphics[width=\linewidth]{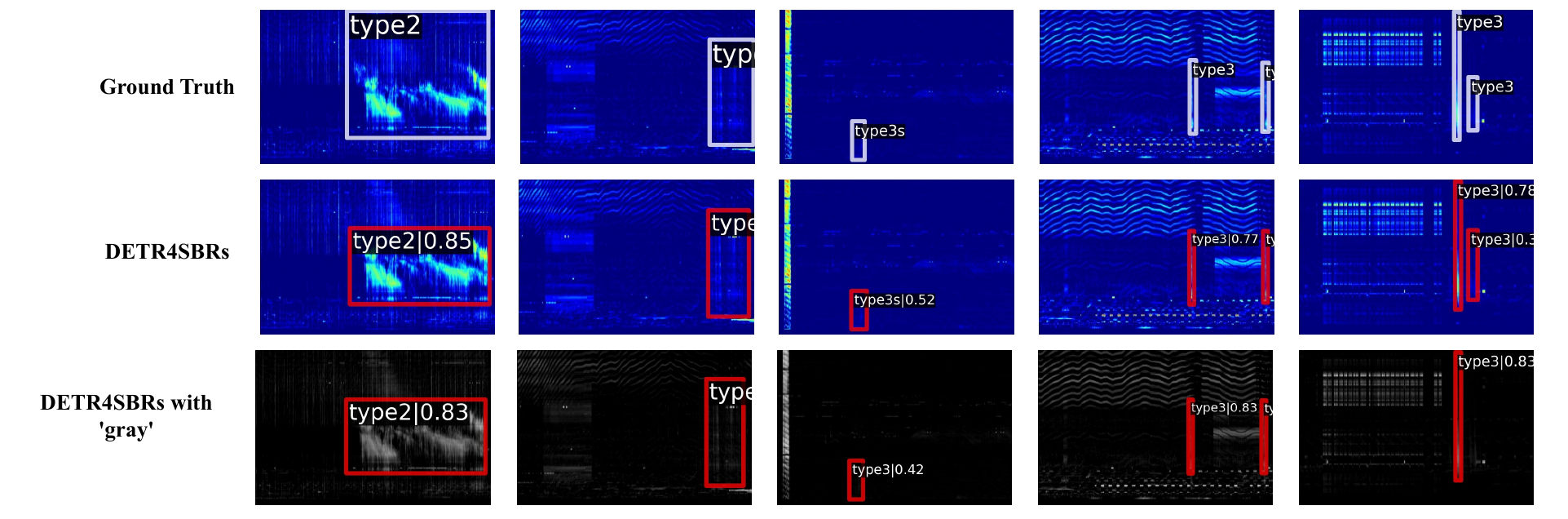}
    \caption{Detection Results of Solar Radio Bursts Under Strong Noise Interference. The first row displays the ground truth boxes (which are manually annotated), while the second row shows the predicted boxes output by DETR4SBRs. It can be observed that, despite the presence of various noise patterns resembling the shape of the eruptions in the spectrogram background, the model is still able to accurately differentiate between the background and the eruption targets. The third row presents the detection results obtained from the grayscale images.}
    \label{fig:fig14}
\end{figure}

\begin{figure}[htb]
    \centering
    \includegraphics[width=\linewidth]{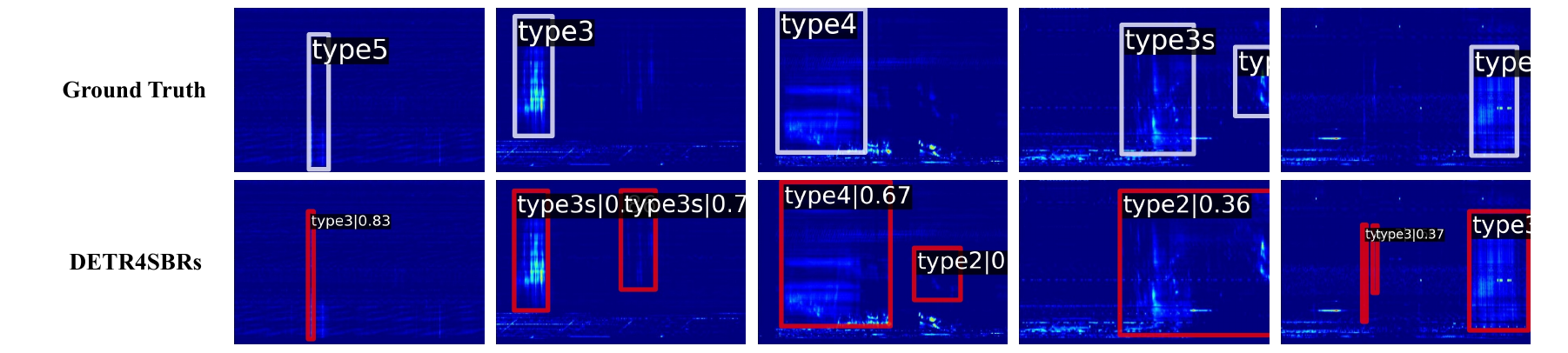}
    \caption{Model Performance in Cases of Mislabeled and Missed Annotations. The first row displays the ground truth boxes (which are manually annotated), while the second row shows the predicted boxes output by DETR4SBRs.The ground truth is manually annotated. It is noteworthy that the model continues to perform effectively even in areas with annotation errors and missed labels. Particularly in the second column, where eruptions are not clearly marked by human annotators, DETR4SBRs is still capable of recognizing these instances.}
    \label{fig:fig15}
\end{figure}
DETR4SBRs, tailored to the characteristics of SRB spectrograms, demonstrates good robustness. The model exhibits high tolerance to noise and can still perform well even in the presence of significant noise in the background, as shown in Figure \ref{fig:fig14}. In Figure \ref{fig:fig14}, the first row shows the annotated ground truth boxes. The second row presents the detection results of DETR4SBRs proposed in this study. The third row refers to the detection results obtained after parsing the FIT files into grayscale images. Although DETR4SBRs was not trained on grayscale images, it still performs effectively.

DETR4SBRs also exhibits good error correction capabilities, as shown in Figure \ref{fig:fig15}. In cases where there are missed annotations or incorrect labels in the manually labeled dataset, DETR4SBRs can correct the mislabeled tags to their correct labels and also detect the missed entries. In Figure \ref{fig:fig15}, the first row shows the annotated ground truth boxes, with some omissions, while the second row shows the detection results of DETR4SBRs proposed in this study, which can detect the SRBs that were missed in the annotations. This indicates that in future practical detection tasks, the model is likely to perform even better than it does on the test set.

\section{Conclusion}

Solar radio bursts (SRBs) detection needs to address the issues of noise and complex scale variations associated with SRBs. This paper proposes a more suitable SSA module to replace the original attention module. Additionally, collaborative hybrid auxiliary training is introduced to address the model's imbalance of positive and negative samples. This model can automatically detect and classify five types of SRBs from many solar radio spectrograms. It can detect different types of SRBs and their locations, achieving a mAP@50 of 83.5$\%$. DETR4SBRs also exhibits high tolerance to noise and accurately regresses the complex scales of SRBs.

SRBs detection requires a high recall rate to minimize missed detections. This capability provides stronger data support for solar physics research and space weather forecasting and warnings. Experimental results demonstrate that DETR4SBRs achieves a recall rate of 99.4$\%$. This means it can efficiently identify most SRBs from the data, significantly reducing the risk of missing important information. The model has shown an exceptional ability to minimize the omission of critical data, providing a solid foundation for further analysis and decision-making. Moreover, the computer vision-based approach we employed requires only a sufficient amount of high-quality annotated image data for training. Theoretically, it can be transferred to the detection of any form of spectrogram images, such as the radio spectra of Earth, Jupiter, and Saturn, or another astronomical image dataset. Researchers interested in this area may utilize DETR4SBRs for further studies. When applied to other datasets, it may be necessary to make some adjustments to DETR4SBRs's modules based on the dataset's characteristics.

However, despite the significant achievements in multiple performance metrics, DETR4SBRs has some key limitations. The most notable issue is the relatively large number of model parameters, leading to slower speeds in practical detection tasks. On the equipment used in this experiment, only about ten SRB spectrogram images can be processed per second. Many parameters mean the model requires more computational resources and time to process data, affecting its efficiency in time-sensitive tasks. Future work can address this issue from two directions. On one hand, more efficient network structures can be explored to reduce the model's computational complexity. This can be achieved by simplifying model parameters or adopting lightweight network designs, thereby improving detection speed. Techniques such as knowledge distillation and network pruning can be employed to reduce unnecessary parameters or explore new network architectures. On the other hand, leveraging hardware acceleration technologies, such as the parallel computing capabilities of GPUs and TPUs, can also alleviate the speed issues caused by many parameters to some extent.

~\\
This research was funded by the National Natural Science Foundation of China (Grant No. 12263008, 62061049), the National Key R\&D Program of China (Grant No.2021YFA1600503), and the Specialized Research Fund for State Key Laboratories of China (Grant No. 202418).

\bibliographystyle{unsrtm}
\bibliography{references}  

\end{document}